\documentclass[a4paper,11pt]{article}
\pdfoutput=1
\usepackage{jcappub}
\usepackage[T1]{fontenc}
\usepackage{multirow}
\usepackage{amsmath} 
\usepackage{amssymb}
\usepackage{mathtools}
\usepackage[version=4]{mhchem}
\usepackage{siunitx}
\usepackage{graphicx}
\usepackage{microtype}

\usepackage{xcolor}
\definecolor{mygreen}{HTML}{04AF70}
\definecolor{myblue}{HTML}{0D98BA}

\usepackage{tikz}
\usetikzlibrary{intersections}
\usetikzlibrary{fadings}
\usetikzlibrary{arrows}

\graphicspath{ {fig/} }

\usepackage{hyperref}
\hypersetup{
    breaklinks,
    colorlinks,
    linktoc=page,
    pdftitle={Echoes from a long time ago: Chewbacca inflation},
    pdfkeywords={CMB, Inflation, Star Wars, Chewbacca},
    pdfauthor={Darth Sidious},
    pdfcreator={\LaTeX}
}

\title{\boldmath Echoes from a long time ago: Chewbacca inflation}

\author[a,b,c,d]{D.~Sidious,}
\author[e,f]{S.~Arcari,}
\author[e,f]{N.~Barbieri,}
\author[e,g]{L.~Bazzanini,}
\author[e,f]{L.~Caloni,}
\author[e,f]{G.~Galloni,}
\author[e,f]{R.~Impavido,}
\author[f]{M.~Lattanzi,}
\author[e,f]{M.~Lembo,}
\author[e,f]{A.~Raffaelli,}
\author[e,f]{N.~Raffuzzi}
\author[e,f,h]{and S.~S.~Sirletti}

\affiliation[a]{Sith Order, Sith Temple, Moraband, Korriban}
\affiliation[b]{Galactic Empire, Imperial Palace, Coruscant}
\affiliation[c]{Galactic Republic, Jedi Temple, Coruscant}
\affiliation[d]{Trade Federation, Cato Neimoidia}
 \affiliation[e]{Dipartimento di Fisica e Scienze della Terra, Università degli Studi di Ferrara, Via G. Saragat 1, I-44122 Ferrara, Italy}
\affiliation[f]{Istituto Nazionale di Fisica Nucleare, Sezione di Ferrara, Via G. Saragat 1, I-44122 Ferrara, Italy}
\affiliation[g]{Istituto Nazionale di Astrofisica - Osservatorio di Astrofisica e Scienza dello Spazio di Bologna, Via Gobetti 101, I-40129 Bologna, Italy}
\affiliation[h]{Dipartimento di Fisica, Università di Trento, Via Sommarive 14, I-38123 Povo, Trento, Italy}

\emailAdd{darth.sidious@empire.gov.so}
\emailAdd{stefano.arcari@unife.it}
\emailAdd{nicola.barbieri@unife.it}
\emailAdd{lorenzo.bazzanini@unife.it}
\emailAdd{luca.caloni@unife.it}
\emailAdd{giacomo.galloni@unife.it}
\emailAdd{riccardo.impavido@unife.it}
\emailAdd{lattanzi@fe.infn.it}
\emailAdd{margherita.lembo@unife.it}
\emailAdd{antonio.raffaelli@unife.it}
\emailAdd{rffnll@unife.it}
\emailAdd{salvatore.sirletti@unitn.it}

\abstract{The cosmic microwave background (CMB) radiation offers a unique avenue for exploring the early Universe's dynamics and evolution. In this paper, we delve into the fascinating realm of slow-roll inflation, contextualizing the primordial acoustic perturbations as the resonant echoes akin to the iconic sound of Chewbacca from the Star Wars universe. By extrapolating polynomial potentials for these primordial sounds, we illuminate their role in shaping the inflationary landscape. Leveraging this framework, we calculate the scalar spectral index ($n_s$) and tensor-to-scalar ratio ($r$), providing insights into the underlying physics governing the inflationary epoch. Employing a rigorous chi-square ($\chi^2$) analysis, we meticulously scrutinize the Planck data combined with that offered by the BICEP/Keck collaboration to identify the Chewbacca sound profile that best aligns with observational constraints. Our findings not only shed light on the intricate interplay between sound and cosmology but also unveil intriguing parallels between the cosmic symphony of the early universe and beloved cultural icons.}

\begin{document}
\maketitle
\flushbottom

\section{Introduction}

The cosmic microwave background (CMB) radiation stands as a cornerstone of modern cosmology, offering a privileged channel for investigating the dynamics and evolution of the early Universe. 
Unveiled by Penzias and Wilson in 1965~\citep{Penzias:1965wn}, the CMB presents an age-old portrait of the universe, offering priceless insights into its fundamental properties and origins.
Anisotropies in the CMB serve as fossilized imprints of primordial perturbations, encoding information about the Universe's initial conditions and subsequent evolution~\citep{Hu:2001bc,Planck:2018nkj}.

In recent decades, the study of inflationary cosmology has emerged as a leading paradigm for explaining the observed large-scale structure of the universe and its remarkable homogeneity and isotropy~\citep{Brout:1977ix,Sato:1981qmu,Guth:1980zm,Starobinsky:1980te,Linde:1981mu,Albrecht:1982wi}. According to the inflationary scenario, the Universe underwent a rapid exponential expansion in its earliest moments, smoothing out any pre-existing irregularities and setting the stage for the subsequent formation of galaxies and large-scale structures. The inflationary paradigm not only elegantly addresses several long-standing problems in cosmology but also makes specific predictions about the statistical properties of primordial perturbations imprinted on the CMB~\citep{Liddle:2000cg}.

``\emph{There is a theory which states that if ever anyone discovers exactly what the Universe is for and why it is here, it will instantly disappear and be replaced by something even more bizarre and inexplicable. There is another theory which states that this has already happened}''~\citep{adams2010restaurant}. 
Inspired by this visionary statement, we delve deeper into the fascinating domain of slow-roll inflation, framing the primordial acoustic perturbations as the resonant echoes akin to the iconic sound of Chewbacca from the Star Wars universe. Just as Chewbacca's distinctive roar reverberates across the galaxy, so too do the primordial perturbations echo through the cosmos, leaving an indelible mark on the fabric of space-time, similar to a periodic Lorentz gas~\citep{Bestiale}.

Our methodology is distinctively unique. We start from the spectrogram of a Chewbacca scream, then perform a Fourier transform to extract its time-integrated spectrum, de-noise it, and finally fit a polynomial potential on top of it. This potential serves as our inflationary model and is used to derive $r$ and $n_s$ for various sounds. We introduce a novel approach for modeling inflationary dynamics by extrapolating polynomial potentials onto the primordial sound data. Leveraging this framework, we calculate the scalar spectral index ($n_s$) and tensor-to-scalar ratio ($r$), unveiling the the underlying physics governing the inflationary epoch.

To validate our methodology and assess its consistency with observational data, we employ a rigorous chi-square ($\chi^2$) analysis. We meticulously exploit the Planck data~\citep{Planck:2018vyg} and BICEP/Keck data~\citep{BICEP:2021xfz} to identify the Chewbacca sound profile that best aligns with observational constraints. Our findings not only shed light on the intricate interplay between sound and cosmology but also unveil intriguing parallels between the cosmic symphony of the early Universe and beloved cultural icons.


\section{Chewbacca-inspired potentials}
\label{sec:potentials}
To explore the intriguing connection between Chewbacca screams and inflationary potentials, we built a full innovative pipeline. We start our procedure with a meticulously selected set of 15 iconic Chewbacca screams\footnote{From now on, each scream will be identified with the diminutive ``Chewie \#''.} sourced from the original Star Wars movies created by George Lucas~\cite{starwarsiv, starwarsv, starwarsvi}. These screams, known for their distinctive timbre and richness in frequency content, serve as our raw data for constructing the inflationary potentials.

\begin{figure}
    \centering
    \includegraphics[width=0.7\textwidth]{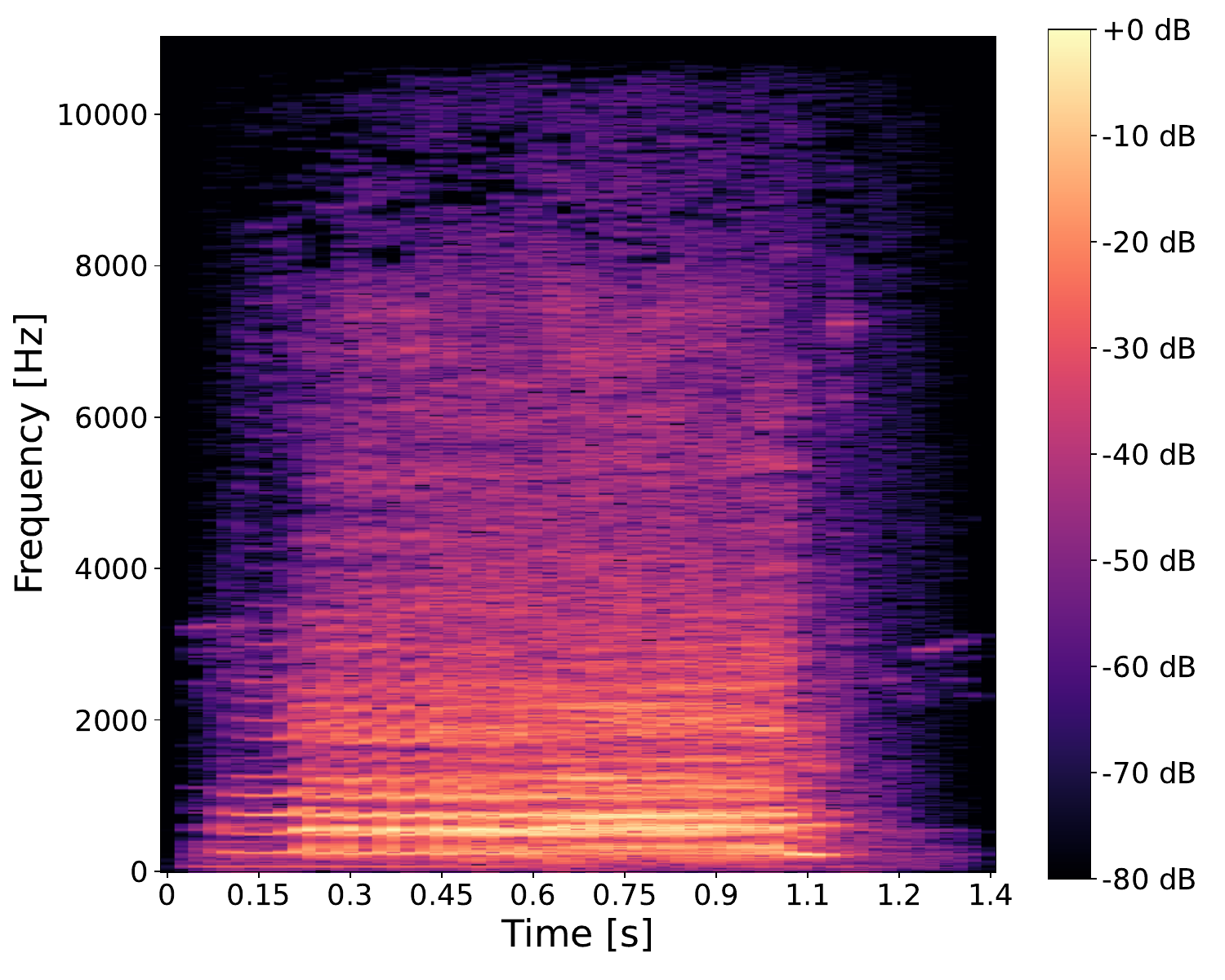}
    \caption{Frequency-time spectrogram obtained through the Fourier analysis of one of fifteen Chewbacca screams from our dataset. The result refers to the sound identifier ``Chewie 10''.}
    \label{fig:spectrogram}
\end{figure}

With the raw audio data in hand, our next step involves the application of Fourier analysis to compute the spectrogram for each Chewbacca scream. The spectrogram, a powerful tool in signal processing, provides a detailed frequency-time representation of the sound signal, revealing its underlying harmonic structure, modulations, and intensity variations~\cite{Allen1977}. Fig.~\ref{fig:spectrogram} shows the spectrogram selected for one of the screams (``Chewie 10''). Notice how most of the information comes from low frequencies, whereas higher frequencies convey noise-related information. To this end, we proceed to enhance the signal quality through a process of de-noising. High-frequency components above 4000 Hz, often associated with noise and artifacts, are systematically removed, ensuring a cleaner representation of the underlying sound signal~\cite{Oppenheim2010}. To further refine the signal and mitigate any remaining fluctuations, we employ smoothing techniques. Top-hat window functions are applied to the de-noised signal, effectively reducing noise and enhancing the stability of the representation~\cite{Bracewell1999}.

With the preprocessed signals at hand, we embark on a journey of mathematical modeling. Leveraging the rich information contained within the smoothed signals, we fit an 8-th degree polynomial function
\begin{equation}
    V(\phi) = \sum_{n=0}^8 a_n\left(\frac{\phi}{M_{\rm Pl}}\right)^n \;,
\end{equation}
to capture the inflationary potential embedded within the Chewbacca screams~\cite{Press1992}. Here, $a_n$ are the polynomial coefficients in units of energy to the fourth and $M_{\rm Pl} = (8\pi G)^{-1/2}\sim2.4\times10^{18}\text{ GeV}$ is the reduced Planck mass.

\begin{figure}
    \centering
    \includegraphics[width=0.7\textwidth]{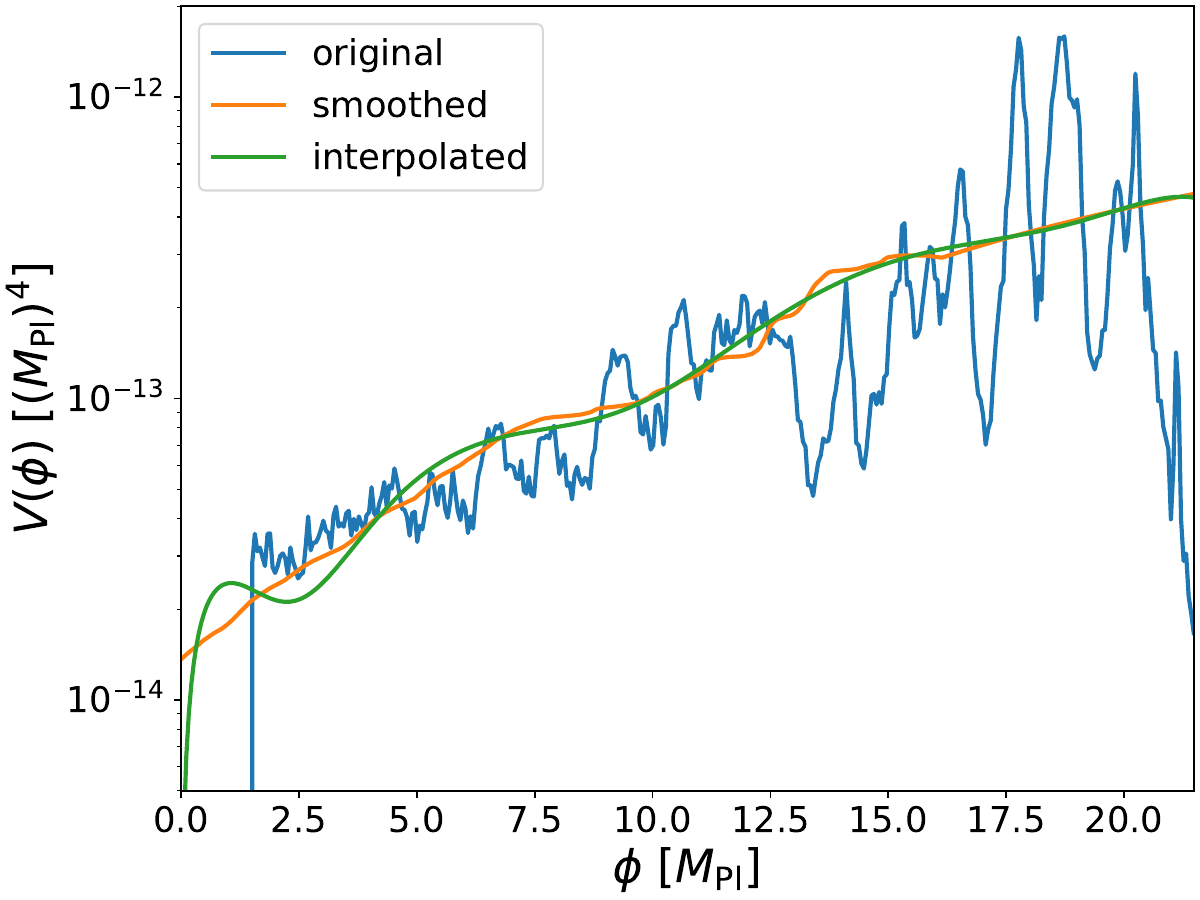}
    \caption{De-noised signal in units of Planck mass given by the time integral of the spectrogram in Fig.~\ref{fig:spectrogram}. The orange line illustrates the smoothed signal, while the green line the 8-th degree polynomial fitted on top of it. The latter serves as our representation of the inflationary potential. The result refers to the scream identifier ``Chewie 10''.}
    \label{fig:pot}
\end{figure}

In our quest to bridge the realms of sound and cosmology, we employ a fundamental unit of measurement: the Planck mass ($M_{\text{Pl}}$). Transforming frequencies into energies in units of Planck mass allows us to express the inflationary potential in a cosmologically meaningful scale, providing insights into the underlying physics of the early Universe. Fig.~\ref{fig:pot} illustrates the underlined procedure by showing the de-noised original signal with the smoothed and fitted signal over-imposed. The result is shown after the transformation to Planck units has been performed for the scream labeled as ``Chewie 10''. Let us notice that the potentials for most of the other screams are similar in shape. In general, the latter is not easily traceable back to standard slow-roll inflation. Nevertheless, it is sufficiently flat in its central part to source an inflationary epoch, ending before the potential begins its descent.

Through this systematic procedure, we derive a set of 15 inflationary potentials, each uniquely inspired by the acoustic characteristics of the corresponding Chewbacca scream. These potentials serve as novel manifestations of the inflationary landscape, shedding light on the intricate interplay between sound phenomena and cosmological dynamics.


\section{Chewbacca inflationary model}
\label{sec:model}
We consider the action of a scalar field minimally coupled to gravity, which is given by
\begin{equation}
    S = \int \mathrm{d}^4x \, \sqrt{-g} \left[ \frac{M_{\rm Pl}^2}{2}R - \frac{1}{2}g^{\mu\nu}\partial_\mu\phi\partial_\nu\phi - V(\phi) \right] \, ,
\end{equation}
where $R$ is the Ricci scalar, $g = \det (g_{\mu\nu})$ is the determinant of the metric tensor and $V(\phi)$ is the potential for the inflaton field. 

We consider 15 inflationary models, defined by the different potentials obtained as described in Sec.~\ref{sec:potentials}. 
From the potential we can define the slow-roll parameters 
\begin{align}
    \epsilon &\equiv \frac{M_{\rm Pl}^2}{2} \left( \frac{V'}{V} \right)^2 \, , \\ 
    \eta &\equiv M_{\rm Pl}^2 \frac{V''}{V} \, , 
\end{align}
which must obey the usual slow-roll conditions, $\epsilon \ll 1$ and $|\eta| \ll 1$.

One of the most appealing features of the inflationary paradigm is that it provides a natural mechanism for giving rise to 
the primordial density fluctuations in the Universe~\citep{Planck:2018jri}. 
Scalar perturbations are commonly characterized in terms of the curvature perturbation, $\zeta$. This is a gauge invariant quantity which is conserved on super-horizon scales~\citep{Bardeen:1983qw, Maldacena:2002vr}. 
In slow-roll inflation, scalar perturbations are characterized by an almost scale-invariant power spectrum, which is commonly parametrized as
\begin{equation}
	\Delta_\zeta(k) = A_s(k_*) \left(\frac{k}{k_*}\right)^{n_s-1} \, ,
\end{equation}
where
\begin{equation}
	\label{eq:scalar-amp}
	A_s(k_*) = \frac{H_*^2}{8\pi^2 M_{\rm Pl}^2 \epsilon_*}
\end{equation}
is the amplitude of scalar fluctuations at the pivot scale $k_*$ and $n_s$ is the scalar spectral index. This is given in terms of the slow-roll parameters by
\begin{equation}
	\label{eq:scalar-tilt}
	n_s - 1 = 2\eta - 6\epsilon \, .
\end{equation} 

Another key prediction of the inflationary paradigm is the production of a stochastic background of primordial gravitational waves, i.e.~transverse and traceless tensor perturbations of the metric~\cite{Guzzetti:2016mkm}.
Analogously to the case of scalar perturbations, the dimensionless power spectrum of primordial tensor modes is almost scale-invariant and can be written in terms of an amplitude $A_t(k_*)$ and a spectral index $n_t$ as
\begin{equation}
	\Delta_t(k) = A_t(k_*) \left(\frac{k}{k_*}\right)^{n_t} \, ,
\end{equation}
where
\begin{equation}
	\label{eq:tensor-amp}
	A_t(k_*) = \frac{2H_*^2}{\pi^2 M_{\rm Pl}^2}
\end{equation}
and
\begin{equation}
	\label{eq:tensor-tilt}
	n_t = -2\epsilon \, .
\end{equation} 
We can then define the tensor-to-scalar ratio
\begin{equation}
	r(k_*)\equiv\frac{A_t(k_*)}{A_s(k_*)} \, ,
\end{equation}
which quantifies the amplitude of tensor perturbations with respect to that of scalar perturbations at the pivot scale $k_*$. In single field slow-roll inflation, the tensor-to-scalar ratio is related to the slow-roll parameter $\epsilon$ by the relation
\begin{equation}
    \label{eq:r}
	r=16\epsilon \, .
\end{equation}

Slow-roll inflationary models are usually characterized in the $r-n_s$ plane. Note that these two quantities are functions of the inflaton field via the slow-roll parameters $\epsilon$ and $\eta$ (see Eqs.~\eqref{eq:scalar-tilt} and~\eqref{eq:r}). In order to compare our theoretical models with observations, we need to evaluate the slow-roll parameters at a given pivot scale of observation, which we choose to be $k_* = 0.05 \; {\rm Mpc}^{-1}$. The value of the field corresponding to the pivot scale $k_*$ can be obtained by requiring that the modes relevant for CMB observations left the horizon $N_{\rm CMB} \sim 60$ e-folds before the end of inflation. This condition can be written as 
\begin{equation}
    \label{eq:efolds}
    N_{\rm CMB} \equiv \int_{t_*}^{t_{\rm end}} H \, \mathrm{d}t \simeq \frac{1}{\sqrt{2}M_{\rm Pl}} \int_{\phi_{\rm end}}^{\phi_*} \frac{\mathrm{d}\phi}{\sqrt{\epsilon(\phi)}}  \, , 
\end{equation}
where $\phi_*$ and $\phi_{\rm end}$ are the values of the field corresponding to the pivot scale $k_*$ and at the end of inflation, respectively.
The latter is fixed by the condition $\epsilon(\phi = \phi_{\rm end}) \simeq 1$. By requiring that $N_{\rm CMB} \in [50,60]$ for $k_* = 0.05 \; {\rm Mpc}^{-1}$, from Eq.~\eqref{eq:efolds} we can derive the value of $\phi_*$.

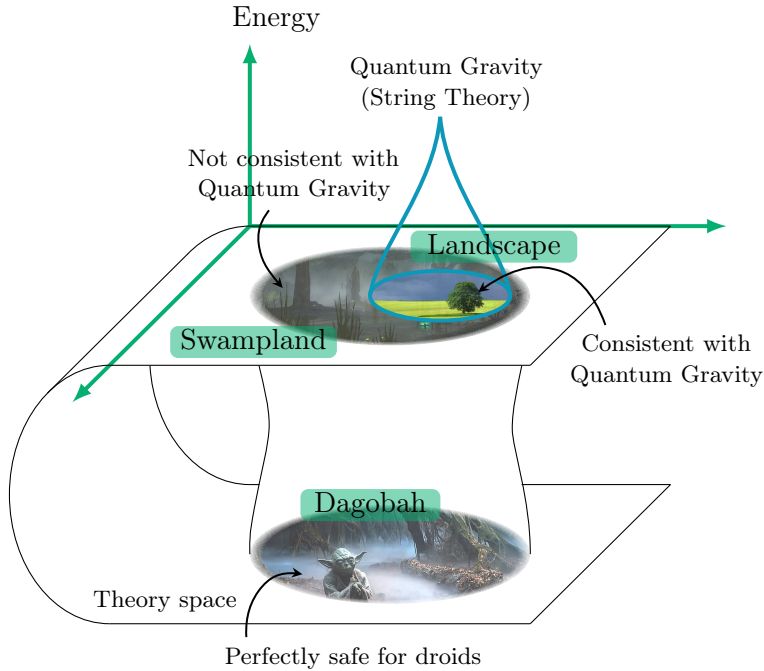
\begin{figure}
    \centering
    \begin{tikzpicture}[x=0.75pt,y=0.75pt,yscale=-1,xscale=1]

\draw [color=mygreen, draw opacity=1] [line width=1.5, -> , >=latex] (260,110) -- (171.41,198.59) ;
\draw [color=mygreen, draw opacity=1] [line width=1.5, -> , >=latex]   (260,110) -- (498,110) ;
\draw [color=mygreen, draw opacity=1] [line width=1.5, <- , >=latex]   (260,18.33) -- (260,110) ;

\draw    (190,310) -- (400,310) ;
\draw    (470,240) -- (400,310) ;

\draw    (395.67,240) -- (470,240) ;

\draw[draw=none] [path picture={\node[opacity=0.8] at (path picture bounding box.center){\includegraphics[width=4cm]{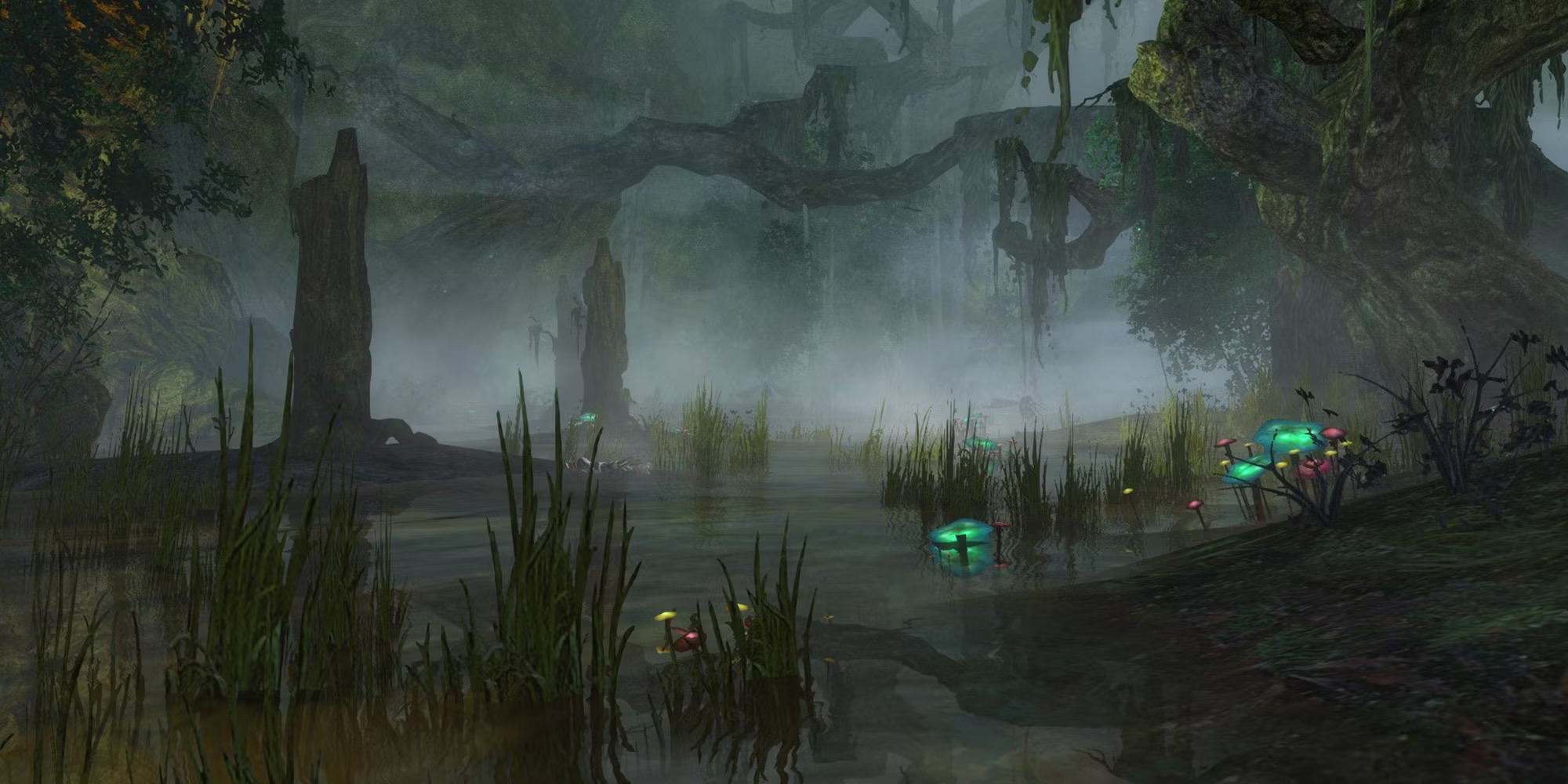}};}] [path fading=circle with fuzzy edge 10 percent] (260,145) .. controls (260,131.19) and (291.34,120) .. (330,120) .. controls (368.66,120) and (400,131.19) .. (400,145) .. controls (400,158.81) and (368.66,170) .. (330,170) .. controls (291.34,170) and (260,158.81) .. (260,145) -- cycle ;

\draw    (190,180) -- (320,180) -- (400,180) ;
\draw    (470,110) -- (400,180) ;

\draw   (190,310) .. controls (190,310) and (190,310) .. (190,310) .. controls (162.39,310) and (140,280.9) .. (140,245) .. controls (140,209.1) and (162.39,180) .. (190,180) ;  
\draw    (229,124) -- (159,194) ;

\draw   (260,240) .. controls (260,240) and (260,240) .. (260,240) .. controls (233.72,240) and (212.17,213.64) .. (210.15,180.14) ;  

\draw   (228.78,124.22) .. controls (237.34,115.32) and (248.19,110) .. (260,110) ;

\draw    (264.58,180.13) .. controls (267.06,190.86) and (269.62,200.41) .. (269.67,210) .. controls (269.71,219.59) and (265.19,235.75) .. (262.31,251.54) .. controls (259.43,267.33) and (260.05,266.39) .. (260,275) ;
\draw    (395.76,180.11) .. controls (393.4,191.75) and (390.96,202.14) .. (391,212) .. controls (391.04,221.86) and (400.17,247.29) .. (400,275) ;
\draw    (260,240) -- (264.75,240) ;
\draw    (260,110) -- (390,110) -- (470,110) ;

\draw  [color=myblue  ,draw opacity=1 ] [path picture={\node[opacity=0.8] at (path picture bounding box.center){\includegraphics[width=2cm]{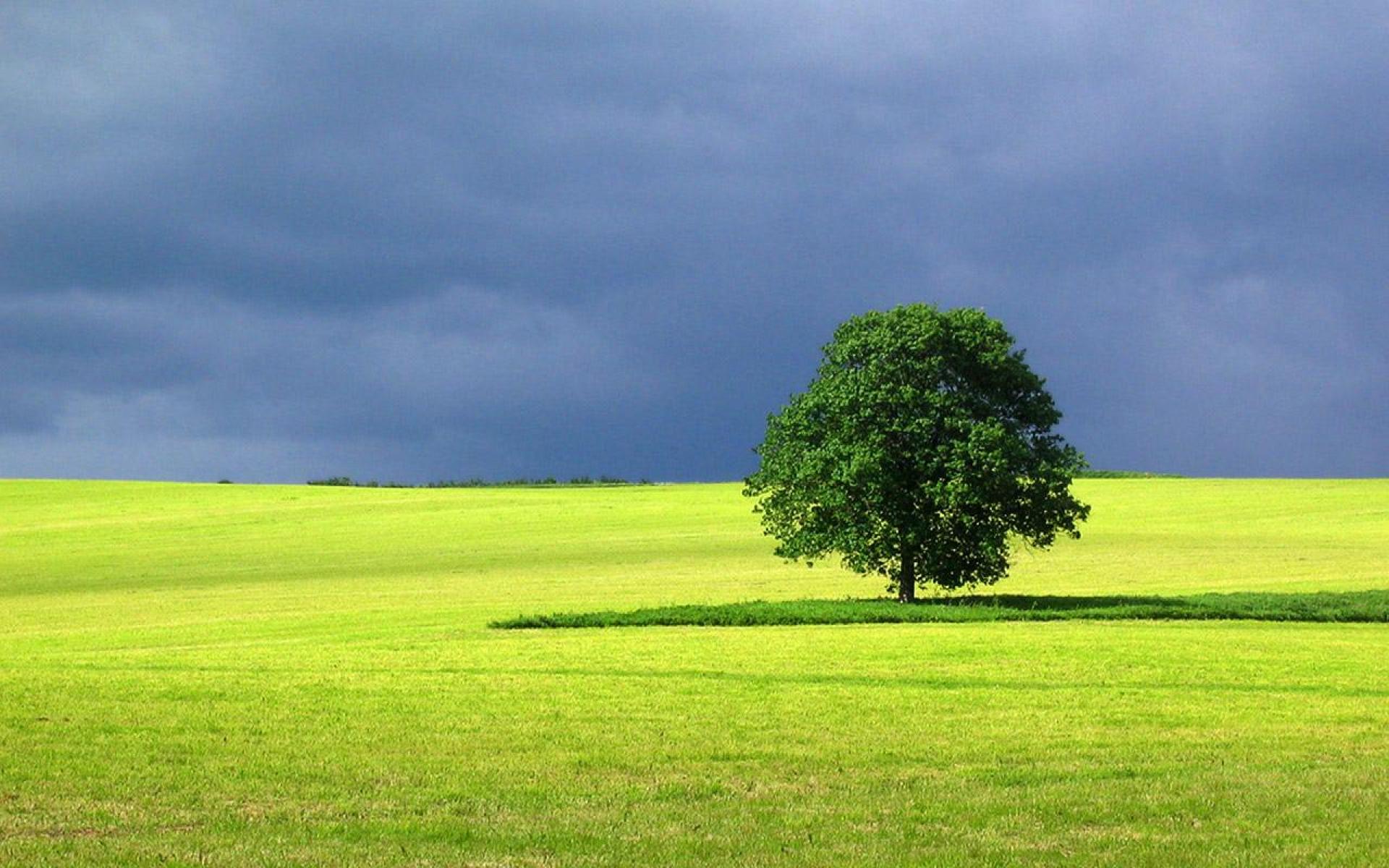}};}] [line width=1.5]  (320,145) .. controls (320,138.1) and (335.67,132.5) .. (355,132.5) .. controls (374.33,132.5) and (390,138.1) .. (390,145) .. controls (390,151.9) and (374.33,157.5) .. (355,157.5) .. controls (335.67,157.5) and (320,151.9) .. (320,145) -- cycle ;
\draw [color=myblue  ,draw opacity=1 ][line width=1.5]    (320,145) .. controls (323,122) and (352,91) .. (355,55) ;
\draw [color=myblue  ,draw opacity=1 ][line width=1.5]    (390,145) .. controls (387,122) and (358,91) .. (355,55) ;

\draw [line width=0.75]    (376.59,141.32) .. controls (394.57,128.39) and (405.88,129.86) .. (414,132.25) .. controls (422.5,134.75) and (431.5,142.75) .. (437.5,159.75) ;
\draw [shift={(374,143.25)}, rotate = 322.43] [fill={rgb, 255:red, 0; green, 0; blue, 0 }  ][line width=0.08]  [draw opacity=0] (5.36,-2.57) -- (0,0) -- (5.36,2.57) -- (3.56,0) -- cycle    ;
\draw [line width=0.75]    (274.44,138.8) .. controls (268.12,131.27) and (265.49,126.99) .. (264.43,118.86) .. controls (263.26,109.92) and (267.86,104.57) .. (269.57,102) ;
\draw [shift={(276.43,141.14)}, rotate = 229.4] [fill={rgb, 255:red, 0; green, 0; blue, 0 }  ][line width=0.08]  [draw opacity=0] (5.36,-2.57) -- (0,0) -- (5.36,2.57) -- (3.56,0) -- cycle    ;

\draw [draw=none] [path picture={\node[opacity=0.8] at (path picture bounding box.center){\includegraphics[width=4cm]{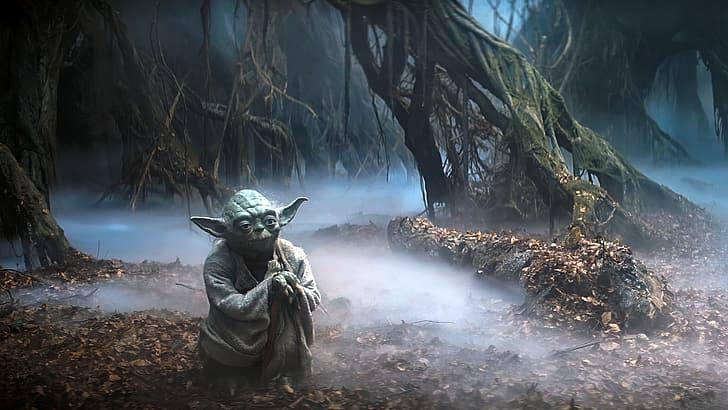}};}] [path fading=circle with fuzzy edge 10 percent] (400,275) .. controls (400,275) and (400,275) .. (400,275) .. controls (400,288.81) and (368.66,300) .. (330,300) .. controls (291.34,300) and (260,288.81) .. (260,275) .. controls (260,261.19) and (291.34,250) .. (330,250) .. controls (368.66,250) and (400,261.19) .. (400,275) -- cycle ;  

\draw [line width=0.75]    (283.19,284.7) .. controls (275.19,283.99) and (269.75,286.16) .. (264.38,293.54) .. controls (258.36,301.83) and (257.91,311.09) .. (258.27,316.73) ;
\draw [shift={(286.27,285.09)}, rotate = 189.06] [fill={rgb, 255:red, 0; green, 0; blue, 0 }  ][line width=0.08]  [draw opacity=0] (5.36,-2.57) -- (0,0) -- (5.36,2.57) -- (3.56,0) -- cycle    ;

\draw (290, 24) node [anchor=north west][inner sep=0.75pt]   [align=left] {\begin{minipage}[lt]{100pt}\setlength\topsep{0pt}
\begin{center}
{ \footnotesize Quantum Gravity \\[-2pt] (String Theory)}
\end{center}

\end{minipage}};
\draw (250, -10) node [anchor=north west][inner sep=0.75pt]   [align=left] {\begin{minipage}[lt]{23.78pt}\setlength\topsep{0pt}
\begin{center}
{Energy}
\end{center}

\end{minipage}};
\draw (150, 292) node [anchor=north west][inner sep=0.75pt]   [align=left] {\begin{minipage}[lt]{100pt}\setlength\topsep{0pt}
\begin{center}
{\footnotesize Theory space}
\end{center}

\end{minipage}};
\draw (400,163.5) node [anchor=north west][inner sep=0.75pt]   [align=left] {\begin{minipage}[lt]{100pt}\setlength\topsep{0pt}
\begin{center}
{\footnotesize Consistent with \\[-2pt]  Quantum Gravity}
\end{center}

\end{minipage}};
\draw (214, 70) node [anchor=north west][inner sep=0.75pt]   [align=left] {\begin{minipage}[lt]{100pt}\setlength\topsep{0pt}
\begin{center}
{\footnotesize Not consistent with \\[-2pt] Quantum Gravity}
\end{center}

\end{minipage}};
\draw (243.67,320.5) node [anchor=north west][inner sep=0.75pt]   [align=left] {\begin{minipage}[lt]{100pt}\setlength\topsep{0pt}
\begin{center}
{\footnotesize Perfectly safe for droids}
\end{center}

\end{minipage}};

\draw (285, 242) node [rectangle,rounded corners=3pt,fill=mygreen,text opacity=1,fill opacity=0.5] [anchor=north west][inner sep=1.2pt]  [align=center] {\begin{minipage}[lt]{50pt}\setlength\topsep{0pt}
\begin{center}
{Dagobah}
\end{center}

\end{minipage}};

\draw (220, 160) node [rectangle,rounded corners=3pt,fill=mygreen,text opacity=1,fill opacity=0.5] [anchor=north west][inner sep=1.2pt]  [align=center] {\begin{minipage}[lt]{60pt}\setlength\topsep{0pt}
\begin{center}
{Swampland}
\end{center}

\end{minipage}};

\draw (340, 112) node [rectangle,rounded corners=3pt,fill=mygreen,text opacity=1,fill opacity=0.5] [anchor=north west][inner sep=1.2pt]  [align=center] {\begin{minipage}[lt]{60pt}\setlength\topsep{0pt}
\begin{center}
{Landscape}
\end{center}

\end{minipage}};
    
    \end{tikzpicture}
    \caption{The Dagobah Swampland.}
    \label{fig:dagobah}
\end{figure}

\section{The Dagobah Swampland}
Gaining a deeper comprehension of the origin of our Universe requires to investigate possible UV completions of the model underlying the inflationary dynamics. Low energy effective theories (EFTs) that cannot be embedded in quantum gravity define the so-called Swampland. This stands in opposition to the string theory landscape, i.e.~the set of EFTs that are compatible with string theory~\citep{Vafa:2005ui}.
Refining the boundaries between the Swampland and the landscape is a highly dynamic field of study, but still far from being completely understood. 
Recently, the so-called Marshland conjecture has been postulated~\citep{Marsh:2019lhu}, marking a crucial step in this direction. 

Here, we argue 
that Chewbacca inflationary models fall into the Swampland.
Given the peculiar origin of Chewbacca-inspired potentials, we introduce a novel region in theory space that we dub \emph{Dagobah Swampland}.
\\
\\
\textbf{The Dagobah Swampland conjecture:} The Dagobah Swampland is a boggy and densely forested region of the swampland, inhospitable for humans but ``perfectly safe for droids'', that can only be reached by jumping into hyperspace via faster-than-light travel (ask Captain Solo for a lift!).
\\
\\
We provide a detailed illustration of the Dagobah Swampland in Fig.~\ref{fig:dagobah}.



\section{Discussion and Conclusions}
\begin{figure}
    \centering
    \includegraphics[width=0.7\textwidth]{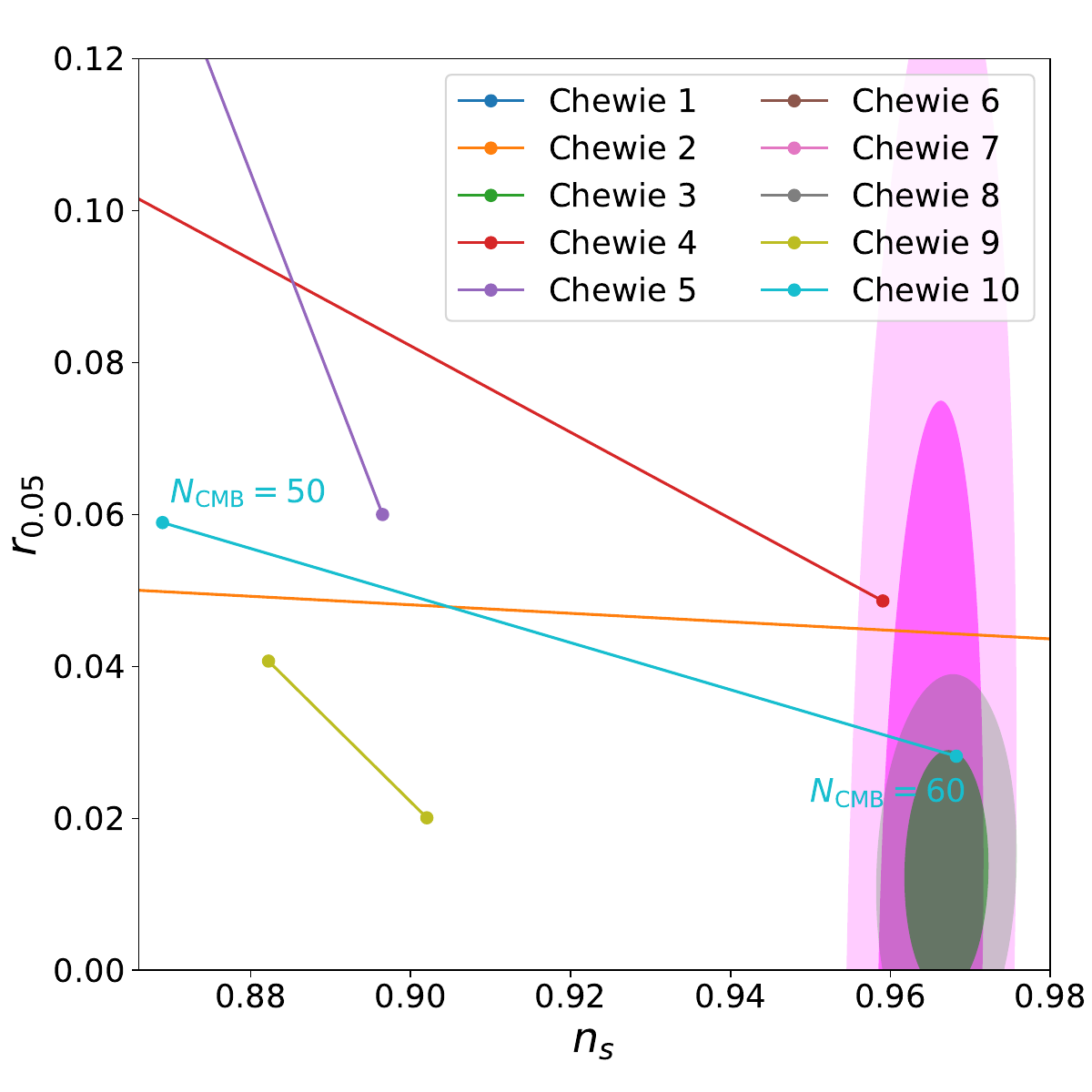}
    \caption{Constraints on Chewbacca inflationary models in the $r-n_s$ plane. The green contours represent the 95\% and 68\% confidence regions obtained from Planck 2018 data~\citep{Planck:2018jri}, while the pink contours are derived from Planck data in combination with BICEP/Keck and BAO data~\citep{BICEP:2021xfz}.}
    \label{fig:constraints}
\end{figure}
Our investigation into the Chewbacca-inspired inflationary models has provided intriguing insights into the potential connections between sound phenomena and cosmological dynamics. Notably, this marks the \emph{debut} of such innovative analysis in literature.
We have systematically analyzed a set of 15 inflationary potentials derived from Chewbacca screams (see Sec.~\ref{sec:potentials}), applying rigorous criteria to assess their viability and compatibility with observational constraints.

Following the procedure detailed in Sec.~\ref{sec:model}, we computed the slow-roll parameters $\epsilon$ and $\eta$ for each inflationary potential as functions of the field $\phi$. By identifying potentials that exhibit sensibly smaller than unity values of $\epsilon$ and $\eta$ at early times, indicative of the onset of slow-roll inflation, we narrowed down our selection to a subset of potentials capable of supporting prolonged inflationary epochs. Furthermore, we scrutinized each potential to determine the presence of a later time at which $\epsilon$ surpasses one, denoting the end of inflation. From our analysis, we found that out of the initial 15 Chewbacca screams, only 10 yielded inflationary scenarios meeting the criteria outlined above. 
\begin{table}[t]
\centering
\begin{tabular}{l|ccc|ccc|}
          & \multicolumn{3}{c|}{$N_{\rm CMB}=50$} & \multicolumn{3}{c|}{$N_{\rm CMB}=60$} \\ \cline{2-7} 
          & $n_s$      & $r$       & $\chi^2$     & $n_s$      & $r$       & $\chi^2$     \\ \hline\hline
Chewie 1  & 1.950      & 0.017     & 43.26        & 0.516      & 1.536     & 112.06       \\
Chewie 2  & 0.783      & 0.055     & 13.88        & 0.999      & 0.043     & 4.53         \\
Chewie 3  & 0.721      & 3e-5      & 23.47        & 0.702      & 0.002     & 17.81        \\
Chewie 4  & 0.853      & 0.109     & 14.12        & 0.959      & 0.049     & 2.76         \\
Chewie 5  & 0.851      & 0.184     & 18.11        & 0.896      & 0.060     & 7.89         \\
Chewie 6  & 1.854      & 0.208     & 97.91        & 0.785      & 0.647     & 39.31        \\
Chewie 7  & 1.495      & 5e-6      & 49.32        & 1.524      & 4e-4      & 51.12        \\
Chewie 8  & 1.768      & 4e-5      & 65.43        & 1.778      & 1e-5      & 67.09        \\
Chewie 9  & 0.882      & 0.041     & 8.22         & 0.902      & 0.020     & 6.48         \\
Chewie 10 & 0.869      & 0.059     & 10.55        & 0.968      & 0.028     & 1.34        
\end{tabular}
\caption{Scalar spectral index ($n_s$) and tensor-to-scalar ratio ($r$) derived from the slow roll parameters of Sec.~\ref{sec:model}, at the pivot field $\phi_*$, for two values of the number of e-folds at CMB. For each value we compute the $\chi^2$ with respect to the observational bounds from Planck18+BICEP/Keck+BAO~\citep{BICEP:2021xfz}. Values below $\chi^2=2$ indicate agreement with the data within 1-$\sigma$.}
\label{tab:res}
\end{table}

Subsequently, we derived the value of the field at which $\epsilon=1$ (denoted as $\phi_{\text{end}}$) and iterated over the integral of the number of e-folds (see Eq.~\eqref{eq:efolds}) to find the pivot value of the field (denoted as $\phi_*$) at which inflation begins. This step ensured a consistent determination of the number of e-folds required for CMB observations. Given the theoretical uncertainty on the related value of $N$, we performed the above computation both for $N_{\rm CMB}=50$ and $N_{\rm CMB}=60$. Using the pivot value of the field, we computed the scalar spectral index ($n_s$) and tensor-to-scalar ratio ($r$) through the related slow-roll parameters at the pivot scale. The results are summarized in Table~\ref{tab:res} and illustrated in Fig.~\ref{fig:constraints} in the $r-n_s$ plane, both for two choices of the number of e-folds at the pivot scale, $N_{\rm CMB} = 50$ and $N_{\rm CMB} = 60$. Let us notice that any viable value between the latter two is represented with a connecting-line in the underlined parameter space.

We investigated our results using a $\chi^2$ analysis to estimate the statistical significance of each scream when compared to the Planck data in combination with BICEP/Keck and BAO data~\citep{BICEP:2021xfz}. The related values obtained from the $\chi^2$ analysis are reported in Table~\ref{tab:res}. Through this analysis, we found that only 3 of the inflationary potentials are compatible with these data. Remarkably, one of these potentials, denoted as ``Chewie 10'', exhibited excellent agreement with observational bounds, aligning with them at less than 1-$\sigma$. This compelling correspondence suggests that ``Chewie 10'' may indeed represent the sound of the Big Bang itself.

In conclusion, our study highlights the potential of utilizing unconventional sources, such as Chewbacca screams, to probe fundamental aspects of cosmology. The identification of ``Chewie 10'' as a plausible candidate for the sound of the Big Bang underscores the intricate interplay between sound phenomena and the underlying dynamics of the early Universe, opening new path for further exploration and investigation.


\acknowledgments
First and foremost, we express our sincere appreciation to George Lucas for creating the iconic character of Chewbacca, whose resonant roar served as the muse for our playful exploration of cosmological phenomena. Without his boundless imagination, this paper would lack the delightful whimsy that characterizes our journey through the cosmos. We reserve a nod of acknowledgment for Frank Herbert, whose seminal work \emph{Dune}~\cite{herbert1965dune} reportedly served as a buffet of inspiration for George Lucas's Star Wars universe. While some may debate the extent of influence, we cannot overlook the echoes of Arrakis in the sands of Tatooine. 
May the Force (and the Spice) be with us all as we continue to unravel the mysteries of the cosmos.\\
We also acknowledge Stronzo Bestiale from the University of Palermo, who was one of the scientists who mostly inspired us to pursue this research. Thanks to his seminal work on the diffusion of Lorentz gasses~\cite{Bestiale}, we were able to link sound waves with inflation paradigm.\\
DS conceived the whole research and couldn't have come up with such a scientific breakthrough without the power of the dark side, after all ``\emph{The dark side of the Force is a pathway to many abilities some consider to be unnatural.}''. Also, he used his Sith mind tricks to force the rest of the authors to perform all the numerical work. This, with the one exception of M.~Lattanzi, who acknowledges signing the paper without doing any actual work (or at least so he thinks), but rather as a well-deserved recognition to his selfless dedication to the continuing mission of educating brave, young minds, to boldly go where no one has gone before. M.~Lattanzi (who is indeed writing these final lines and finds very funny to speak of himself in the third person) would also like to file a minority report, and, notwithstanding his love for the Star Wars and Dune sagas, complain about the absence of any Star Trek reference or joke in the paper. It would have been as easy as citing some (serious) research paper on the Cardassian expansion~\citep{Freese:2002sq}, or on the Q-continuum cosmological N-body simulations~\citep{Heitmann:2014dba}. He would definitely not let the paper in this form go through review, if he was the referee. 



\bibliographystyle{JHEP}
\bibliography{ref}

\end{document}